\documentclass[preprint,preprintnumbers,nofootinbib,aps,10pt,twocolumn]{revtex4-1}
\usepackage{amsmath,amssymb,bm,epsfig}
\usepackage{color}
\usepackage{natbib}
\usepackage{hyperref} 
\usepackage{ulem}
\usepackage{graphicx}
\usepackage{xcolor,colortbl}
\RequirePackage{lineno}
\newcommand{\nn}{\nonumber}
\newcommand{\sNN}{\sqrt{s_{\textrm{NN}}}}

\definecolor{Gray}{gray}{0.85}
\newcolumntype{a}{>{\columncolor{Gray}}c}

\def \beq{\begin{equation}}
\def \eeq{\end{equation}}
\def \beqa{\begin{eqnarray}}
\def \eeqa{\end{eqnarray}}
\def \la{\langle}
\def \ra{\rangle}

\begin{document}

\title{Splitting of elliptic flow in a tilted fireball}

\author{Tribhuban Parida}
\email{tribhubanp18@iiserbpr.ac.in}
\author{Sandeep Chatterjee}
\email{sandeep@iiserbpr.ac.in}

\affiliation{Department of Physical Sciences,\\
Indian Institute of Science Education and Research Berhampur,\\ 
Transit Campus (Govt ITI), Berhampur-760010, Odisha, India}

\begin{abstract}

The splitting of elliptic flow measured in different regions of the momentum space of produced 
hadrons has been recently studied in transport models and proposed as a sensitive probe of the 
angular momentum carried by the fireball produced in a relativistic heavy ion collision. The initial 
state angular momentum also gives rise to rapidity odd directed flow which has been measured. 
We consider a relativistic hydrodynamic framework with the initial matter distribution suitably 
calibrated to describe the observed directed flow and apply it to study the spilt in the elliptic flow. 
Our study suggests that the split in the elliptic flow is mostly driven by directed and triangular 
flows and may be used to constrain models of initial state rapidity distribution of matter in the 
fireball.

\end{abstract}

\maketitle

\section{Introduction}

The system of two non-central colliding relativistic heavy ion nuclei carries large angular 
momentum. In the aftermath of the collision, a part of this angular momentum is deposited in 
the locally thermalised fireball. The hydrodynamic response results in fluid vorticity and possibly 
spin polarisation which may be finally observed in the phase space occupation and polarisation 
states of the emitted particles~\cite{Liang:2004ph,Liang:2004xn,Becattini:2007sr,Betz:2007kg,
Ipp:2007ng,Becattini:2013vja,Becattini:2015ska,Pang:2016igs,Karpenko:2016jyx}.

There have been several attempts to model the initial longitudinal distribution of various 
hydrodynamic fields like the energy density and fluid velocity which after hydrodynamic 
evolution and particlization can leave their imprint on different observables~\cite{Snellings:1999bt,
Betz:2007kg,Becattini:2013vja,Becattini:2015ska,Pang:2016igs,Karpenko:2016jyx,Bozek:2010bi,
Bozek:2010vz,Bozek:2015bha,Bozek:2015bna,Bozek:2015tca,Broniowski:2015oif,
Chatterjee:2017mhc,Bozek:2017qir,Pang:2015zrq,Pang:2014pxa,Pang:2018zzo,Wu:2018cpc,
Shen:2020jwv,Ryu:2021lnx,Jiang:2021foj,Jiang:2021ajc}. While for quite some time the rapidity 
dependence of directed flow has been used to discriminate such models of the initial three 
dimensional matter distribution~\cite{Snellings:1999bt,Bozek:2010bi,Shen:2020jwv,Ryu:2021lnx,
Jiang:2021foj,Jiang:2021ajc}, recently it has been demonstrated that even polarisation 
measurements of the final state hadrons can constrain such ansatz of initial matter 
distribution~\cite{Ryu:2021lnx,Alzhrani:2022dpi}.

It has been pointed out that the non-zero angular momentum of the fireball results in the 
splitting of elliptic flow in the momentum space of the final state hadrons~\cite{Chen:2021wiv}. 
This makes the splitting of elliptic flow a sensitive observable to constrain models of the initial 
state three dimensional matter distribution of the fireball. In a subsequent work, it was argued that 
this splitting is mainly driven by the directed flow~\cite{Zhang:2021cjt}. These studies were 
conducted in a transport model framework and they don't describe the data on directed 
flow~\cite{STAR:2008jgm}. This raises concern on their model prediction of the 
elliptic flow splitting.

The various models of the longitudinal profile of the fireball  that has been explored so far can be 
broadly divided into two categories: shifted initial condition (SIC)~\cite{Hirano:2002ds} and tilted initial 
condition (TIC)~\cite{Bozek:2010bi}. In SIC, the rapidity profile at each point on the transverse plane is 
shifted according to the local centre of mass rapidity. TIC is inspired from the ansatz that a participant nucleon 
deposits more energy along its direction of motion~\cite{Brodsky:1977de,Back:2002wb,Bialas:2004su,
Adil:2005qn,Armesto:2006bv,Bzdak:2009dr}. It has been shown that SIC fails to describe the 
rapidity slope of directed flow at mid-rapidity in Au+Au at $\sNN=200$ GeV while TIC succeeds to 
describe the experimental data on directed flow, albeit with a free parameter $\eta_m$ that 
parametrises the forward-backward asymmetry in the energy deposition of a participant in the initial 
state~\cite{Bozek:2010bi}. In this study, we discuss the contrasting nature of the splitting of the elliptic 
flow in different regions of the momentum space for both SIC as well as TIC, underlying the significance of 
this observable in our efforts to comprehend the longitudinal dynamics of the fireball.

\section{INITIAL RAPIDITY PROFILE}

We study Au+Au collisions at $\sNN=200$ GeV. The initial condition for the hydrodynamic evolution 
of the fireball is obtained from the Optical Glauber model. Here, the nucleus is modelled as a 
Woods-Saxon distribution $ \rho(x,y,z) = \frac{\rho_0} {1 + \exp{ ( \frac{r-R}{a}}  ) }$ where 
$\ r = \sqrt{x^2 + y^2 + z^2 }$, $R = 6.38$ fm and $a = 0.535$ fm~\cite{Shou:2014eya}. The z-axis 
is taken along the beam axis while the x-axis is along the impact parameter direction. The nuclear 
thickness function $T(x, y)$ is obtained as
\begin{equation}
    T(x,y) = \int \rho(x,y,z) dz  
\end{equation}
using which one can define the total forward ($N_+$) and backward ($N_-$) going participants along the 
beam axis at a point $(x,y)$ on the transverse plane
\begin{equation}
    N_{+} (x,y) = T(x - b/2,y ) \left( 1 - (1 - \sigma_{NN} T(x+b/2,y))\right) 
\end{equation}
\begin{equation}
    N_{-} (x,y) = T(x + b/2,y ) \left( 1 - (1 - \sigma_{NN} T(x-b/2,y))\right) 
\end{equation}

We have compared between two models of $\epsilon(x,y,\eta_s)$, the initial energy density deposited at a 
constant $\tau$ hypersurface at $(x,y,\eta_s)$: SIC~\cite{Hirano:2002ds} and TIC~\cite{Bozek:2010bi}. 

In case of SIC, the following ansatz is adopted for $\epsilon(x,y,\eta_s)$
\beqa
  \epsilon(x,y,\eta_{s}) &=& \epsilon_{0} \left[ \left( N_{+}(x,y) + N_{-}(x,y)  \right)  \frac{ (1- \alpha)}{2} \right. \nn\\
  &&\left. +  N_{coll} (x,y) \alpha\right] \times \epsilon_{\eta_s}(\eta_s-\eta_{sh}(x,y))
%  &&\times f\left( \eta_{s} - \eta_{sh} \right)  
\label{eq.shift}
\eeqa
where $\epsilon_{\eta_s}(\eta_s-\eta_{sh}(x,y))$ gives the $\eta_s$ distribution at $(x,y)$
\begin{equation}
  \epsilon_{\eta_s}(\eta_s) = \exp \left(  -\frac{ \left( \vert \eta_{s} \vert - \eta_{0} \right)^2}{2 \sigma_{\eta}^2}   
    \theta (\vert \eta_{s} \vert - \eta_{0} ) \right)
    \label{eq.seven}
\end{equation}
We have used $\epsilon_0 = 13.2~\text{GeV/fm}^3$, $\alpha=0.14$, $\eta_{0}=1.3$ and $\sigma_{\eta} = 1.5$ that provides a 
good description of the ($\eta-\frac{dN_{ch}}{d\eta}$) data. $\eta_{sh}(x,y)$ is given by
\begin{equation}
    \eta_{sh} = \frac{1}{2} \ln \frac{N_+(x,y) + N_-(x,y) + v_N (N_+(x,y) - N_-(x,y) ) }{N_+(x,y) + 
    N_-(x,y) - v_N (N_+(x,y) - N_-(x,y) )}
\end{equation}
Here, $v_N$ is the initial longitudinal velocity of each nucleon with mass $m_N$.
\begin{equation}
    v_{N} = \sqrt{1 - (4m_{N}^2) / (s_{NN})}
\end{equation}

The second initial condition that we have studied is the TIC. In this case $\epsilon(x,y,\eta_{s})$ 
is given by

\beqa
  \epsilon(x,y,\eta_{s}) &=& \epsilon_{0} \left[ \left( N_{+}(x,y) f_{+}(\eta_{s}) + N_{-}(x,y) f_{-}(\eta_{s})  \right)\right.\nn\\
                           &&\left.\times \left( 1- \alpha \right) + N_{coll} (x,y)  \epsilon_{\eta_s}\left(\eta_{s}\right) \alpha \right] 
 \label{eq.tilt}
\eeqa
where $\epsilon_{\eta_s}(\eta_s)$ is the rapidity even profile as given in Eq.~\ref{eq.seven} and $ f_{+,-}(\eta_s)$ 
introduce rapidity odd component in $\epsilon$ 
\begin{equation}
    f_{+,-}(\eta_s) = \epsilon_{\eta_s}(\eta_s) \epsilon_{F,B}(\eta_s)
\end{equation}
where
\begin{equation}
    \epsilon_{F}(\eta_s) = 
    \begin{cases}
    0, & \text{if } \eta_{s} < -\eta_{m}\\
    \frac{\eta_{s} + \eta_{m }}{2 \eta_{m}},  & \text{if }  -\eta_{m} \le \eta_{s} \le \eta_{m} \\
    1,& \text{if }  \eta_{m} < \eta_{s}
\end{cases}
\end{equation}
and 
\begin{equation}
    \epsilon_{B} (\eta_s) = \epsilon_F(-\eta_s)
\end{equation}
We have used $\eta_{m} = 2.5$ to describe the directed flow data~\cite{STAR:2008jgm}.
For both the initial conditions, we have assumed the Bjorken flow ansatz 
\begin{equation*}
u^{\mu} (\tau_{0},x,y,\eta_s) = (\cosh {\eta_s}, 0, 0, \sinh{\eta_s} ) 
\end{equation*} 

We evolve the above deposited initial energy distribution with the publicly available MUSIC code~\cite{Schenke:2010nt,
Schenke:2011bn,Denicol:2018wdp,Paquet:2015lta} which implements evolution within the framework of 3+1 D 
relativistic hydrodynamics followed by Cooper Frye freezeout at $T=150$ MeV and finally allowing all the resonances to 
decay to stable hadrons under strong interaction. Thus, we obtain the momentum space probability distribution of 
hadrons using which we compute various observables. We consider the lattice QCD based equation of state, 
NEoS-B at zero baryon density~\cite{Monnai:2019hkn,HotQCD:2014kol,HotQCD:2012fhj,Ding:2015fca} and 
take the shear viscosity $\eta$ to entropy density $s$ ratio, $\eta/s=0.08$. We have ignored the effects of bulk viscosity.

\section{SPLITTING OF THE ELLIPTIC FLOW}

The azimuthal distribution of the hadrons in the plane transverse to the beam axis can be expanded into Fourier 
components in the following way:
\beq
\frac{dN}{d\phi} = \frac{1}{2\pi}\left(1 + 2 \sum_n \left(v_n \cos(n(\phi-\psi_\text{RP})) + s_n \sin(n(\phi-\psi_\text{RP}))\right) \right)
\eeq
where $\psi_{\text{RP}}$ is the reaction plane angle in the laboratory frame. $v_n$ and $s_n$ are the Fourier coefficients 
that characterise the distribution.

There has been a recent proposal to measure the split $\Delta v_2$ 
in $v_2 = \la \cos(2(\phi-\psi_\text{RP})) \ra$, as measured in different regions of the final hadron momentum space
\beq
\Delta v_2 = {v_2}^{\text{R}} - {v_2}^{\text{L}}
\label{eq.deltav2}
\eeq
where ${v_2}^{\text{R}} = \la \cos(2(\phi^{\text{R}}-\psi_\text{RP})) \ra$ with $\phi^{\text{R}}\in((\psi_{\text{RP}}-\pi/2), 
(\psi_{\text{RP}}+\pi/2))$ and ${v_2}^{\text{L}} = \la \cos(2(\phi^{\text{L}}-\psi_\text{RP})) \ra$ with 
$\phi^{\text{L}}\in((\psi_{\text{RP}}+\pi/2), (\psi_{\text{RP}}+3\pi/2))$. Here, $\la...\ra$ refers to averaging over 
the phase space of the produced hadrons.

$\psi_\text{RP}$ is not directly measurable in experiments. The second order event plane orientation $\psi_2$ 
and the first order spectator plane $\psi_\text{SP}$ have been proposed as good proxies for 
$\psi_\text{RP}$~\cite{STAR:2013ksd,ALICE:2012nhw}. 
However, for the determination of $\Delta v_2$, $\psi_\text{SP}$ alone is suitable as $\psi_{2}=\pi$ is identified with $\psi_{2}=0$ 
and hence does not distinguish between the phase spaces associated with $\phi^{\text{R}}$ and $\phi^{\text{L}}$. Recently, the 
Event Plane Detector has been installed at large rapidities which can also be used to estimate 
$\psi_\text{RP}$\cite{Adams:2019fpo}.

${v_2}^{\text{R}}$ and ${v_2}^{\text{L}}$ work out to be the following~\cite{Zhang:2021cjt}:
\begin{eqnarray}
     v_{2}^{\text{R}} &=& 
    \frac{\int_{\psi_\text{RP}-\frac{\pi}{2}}^{\psi_\text{RP}+\frac{\pi}{2}} \ \cos(2(\phi-\psi_\text{RP})) \frac{dN}{d\phi} \  d \phi } {\int_{\psi_\text{RP}-\frac{\pi}{2}}^{\psi_\text{RP}+\frac{\pi}{2}}  \frac{dN}{d\phi} d \phi} \nn\\
    &\approx& 
    \frac{v_{2}  + \frac{4 v_{1}}{3 \pi} + \frac{12 v_3}{5 \pi} - \frac{20v_5}{21 \pi}  }{ 1 + \frac{4 v_1}{\pi} - \frac{4 v_3}{3 \pi} + 
    \frac{4 v_5}{5 \pi} }
    \label{eq.v2r}
\end{eqnarray}

\begin{eqnarray}
v_{2}^{\text{L}} &=& 
    \frac{\int_{\psi_\text{RP}+\frac{\pi}{2}}^{\psi_\text{RP}+\frac{3 \pi}{2}} \ \cos(2(\phi-\psi_\text{RP})) \frac{dN}{d\phi} \  d \phi } {\int_{\psi_\text{RP}+\frac{\pi}{2}}^{\psi_\text{RP}+\frac{3 \pi}{2}}  \frac{dN}{d\phi}  d \phi} \nn\\
    &\approx& 
    \frac{v_{2}  - \frac{4 v_{1}}{3 \pi} - \frac{12 v_3}{5 \pi} + \frac{20v_5}{21 \pi}  }{ 1 - \frac{4 v_1}{\pi} + \frac{4 v_3}{3 \pi} - 
    \frac{4 v_5}{5 \pi} }
    \label{eq.v2l}
\end{eqnarray}

In Eqs.~\ref{eq.v2r} and \ref{eq.v2l} we have omitted contributions from harmonics higher than the fifth order. Further, 
from Eqs.~\ref{eq.deltav2}, \ref{eq.v2r} and \ref{eq.v2l} we get
\begin{equation}
\Delta v_2 \approx \frac{8v_1}{3\pi} + \frac{24v_3}{5\pi} - \frac{40v_5}{21\pi}
\label{eq.deltav2ap}
\end{equation}

In Eq.~\ref{eq.deltav2ap} only terms which are linear in the flow harmonics have been shown as they are sufficient to 
estimate $\Delta v_2$ . Thus, $\Delta v_2$ is sourced mainly by the odd flow harmonics. The collision geometry is such that 
these odd flow harmonics have rapidity odd components with respect to the reaction plane. Hence, $\Delta v_2$ is also rapidity 
odd following the odd flow harmonics which is also evident from Eq.~\ref{eq.deltav2}. There has been measurement of directed 
flow with respect to $\psi_\text{SP}$~\cite{STAR:2008jgm}. Optical Glauber model with TIC followed by hydrodynamic expansion 
is able to describe the $v_1$ measurement at $\sNN=200$ GeV for Au+Au collisions\cite{Bozek:2010bi}. Here, we use similar TIC 
within an optical Glauber model followed by hydrodynamic expansion to compute the model expectation for $\Delta v_2$. We 
expect similar results on including fluctuations in the initial condition~\cite{Bozek:2011ua}.

\section{RESULTS}

\begin{figure}
 \begin{center}
 \includegraphics[scale=0.25]{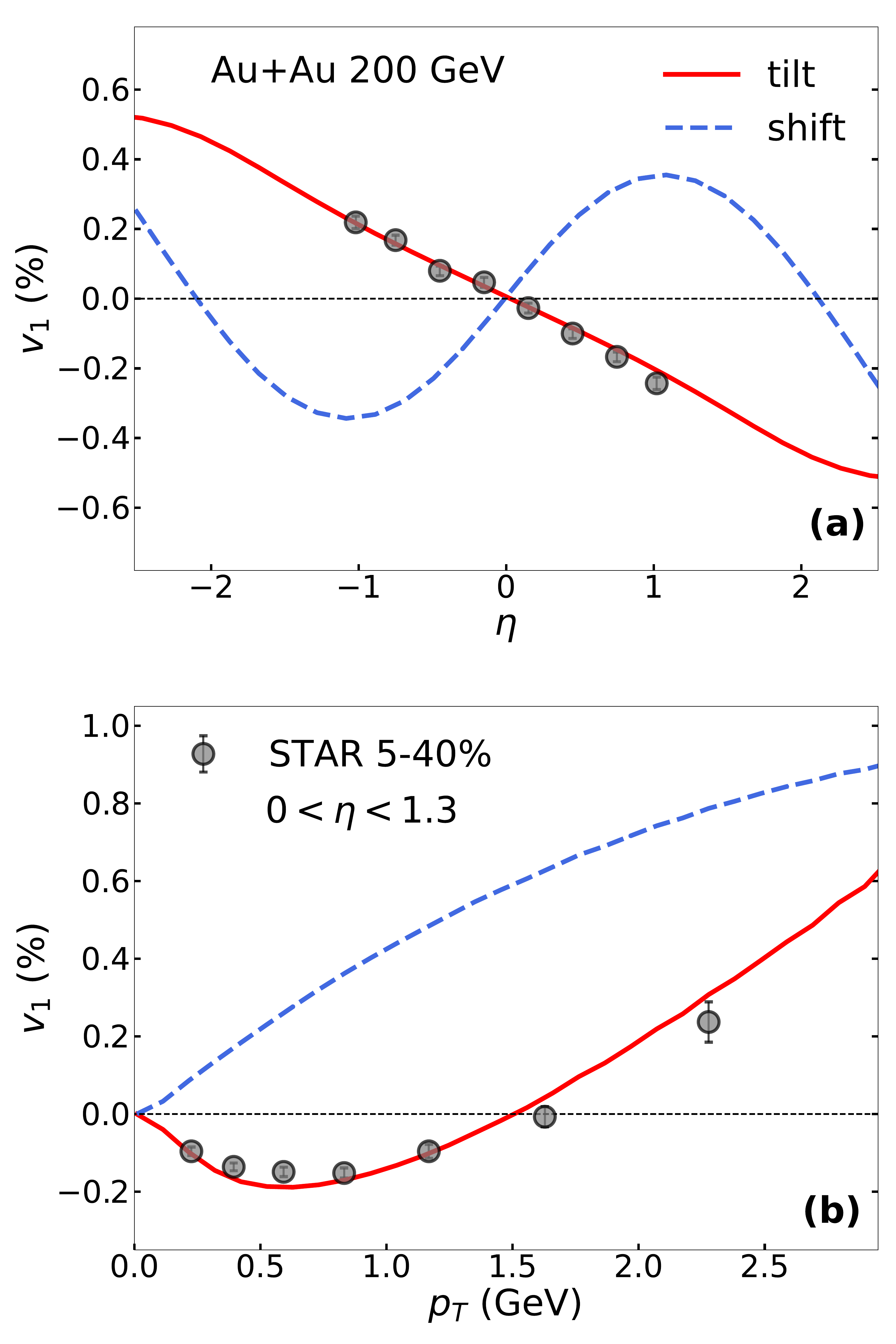}
 \caption{(Color online) Phase space dependence of $v_1$ is computed both for tilted initial condition (red solid line) and 
 shifted initial condition (blue dashed line) for  5-40 $\%$ centrality Au+Au collisions at $\sNN=200$ GeV. The model 
 expectations are compared to measurements from the STAR collaboration~\cite{STAR:2008jgm}. $v_1$ vs $\eta$ is shown in 
 panel (a) and $v_1$ vs $p_T$ is shown in panel (b).}
 \label{fig.v1}
 \end{center}
\end{figure}

We will now present the prediction for $\Delta v_2$ with respect to the reaction plane as computed with TIC as well as SIC. 
As seen in Eq.~\ref{eq.deltav2ap}, the leading contributions to $\Delta v_2$ arise from the odd harmonics, out of which there 
are STAR measurements on $v_1$ with respect to the spectator plane~\cite{STAR:2008jgm}. Thus, we first compare the model 
results with the STAR data for $v_1$  in Fig.~\ref{fig.v1}. We have plotted the model expectations for $v_1-\eta$ and $v_1-p_T$ 
in the panels (a) and (b) respectively and compared them 
to the STAR measurements~\cite{STAR:2008jgm}. The model results have been shown for both TIC (red 
solid line) as well as those from SIC (dashed blue line). We find that for both $v_1-\eta$ and $v_1-p_T$, 
TIC is able to describe the STAR data well while SIC fails. This is in agreement with earlier 
studies~\cite{Bozek:2010bi,Jiang:2021ajc}. Thus, we expect TIC to provide correct prediction of $\Delta v_2$.

\begin{figure}
 \begin{center}
 \includegraphics[scale=0.25]{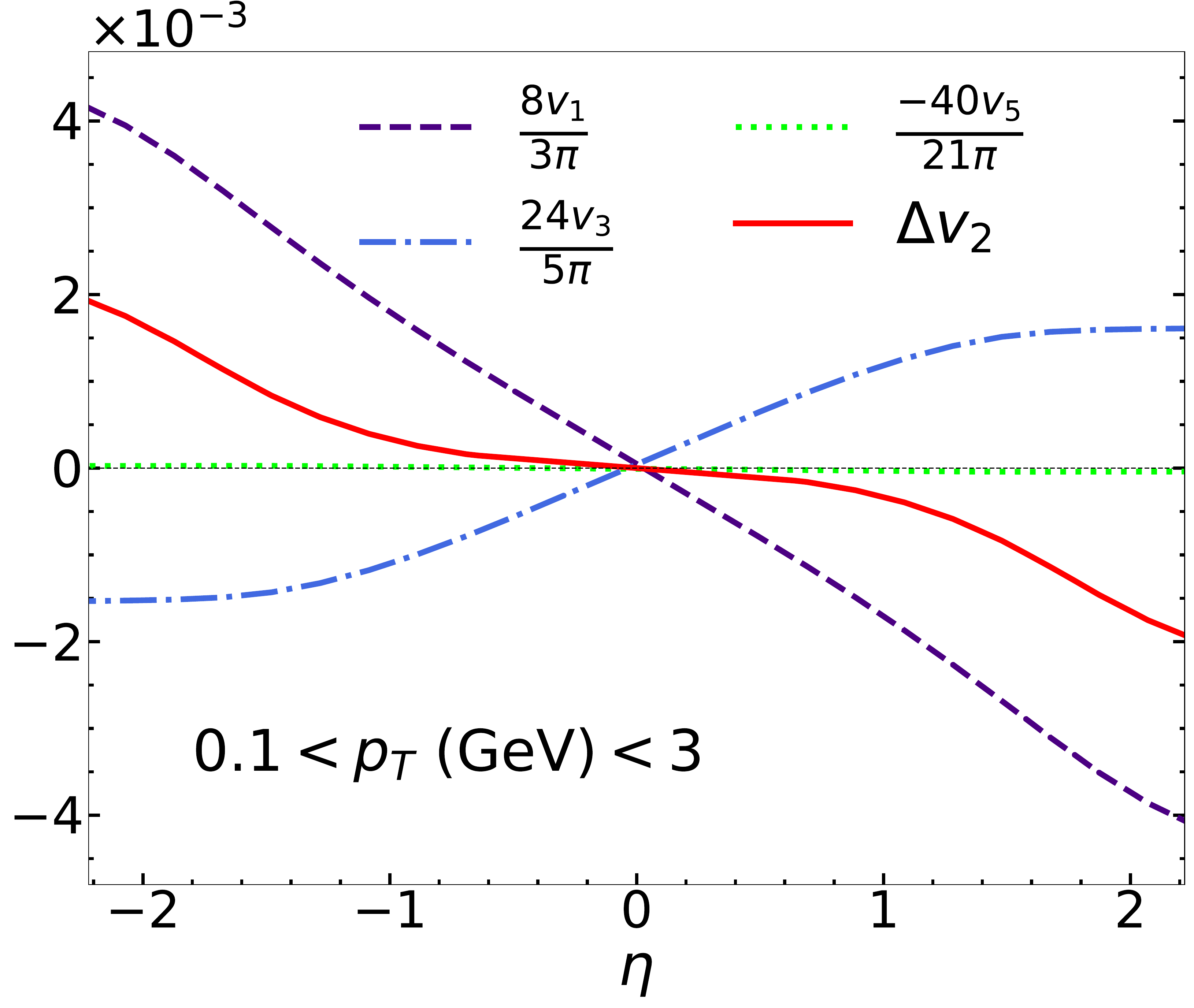}
 \caption{(Color online) The prediction for $\Delta v_2$ vs $\eta$ with tilt initial condition has been plotted in red solid line. 
 Further, the first three leading flow harmonics that contribute to $\Delta v_2$ (see Eq.~\ref{eq.deltav2ap}) are shown as well.}
 \label{fig.deltav2eta}
 \end{center}
\end{figure}

The good description of $v_1$ by TIC motivates us further to compute $\Delta v_2$ vs $\eta$ 
within the same scheme. In Fig.~\ref{fig.deltav2eta} we have shown the model predictions for the $\eta$ dependence 
of $\Delta v_2$. Further, we have also plotted the odd harmonics along with appropriate coefficients as suggested by 
Eq.~\ref{eq.deltav2ap} that are the dominant contributors to $\Delta v_2$. Firstly, we note that $\Delta v_2$ arises as a 
competition between $\frac{8v_1}{3\pi}$ and $\frac{24v_3}{5\pi}$ as they are of opposite signs. The $\frac{8v_1}{3\pi}$ term 
marginally wins and hence $\Delta v_2$ follows its sign. For $\vert \eta \vert <1$, $\Delta v_2\sim10^{-4}$ and it grows substantially 
for larger $\eta$. The contribution from $v_5$ is negligible in the entire $\eta$ range. 

\begin{figure}
 \begin{center}
  \includegraphics[scale=0.25]{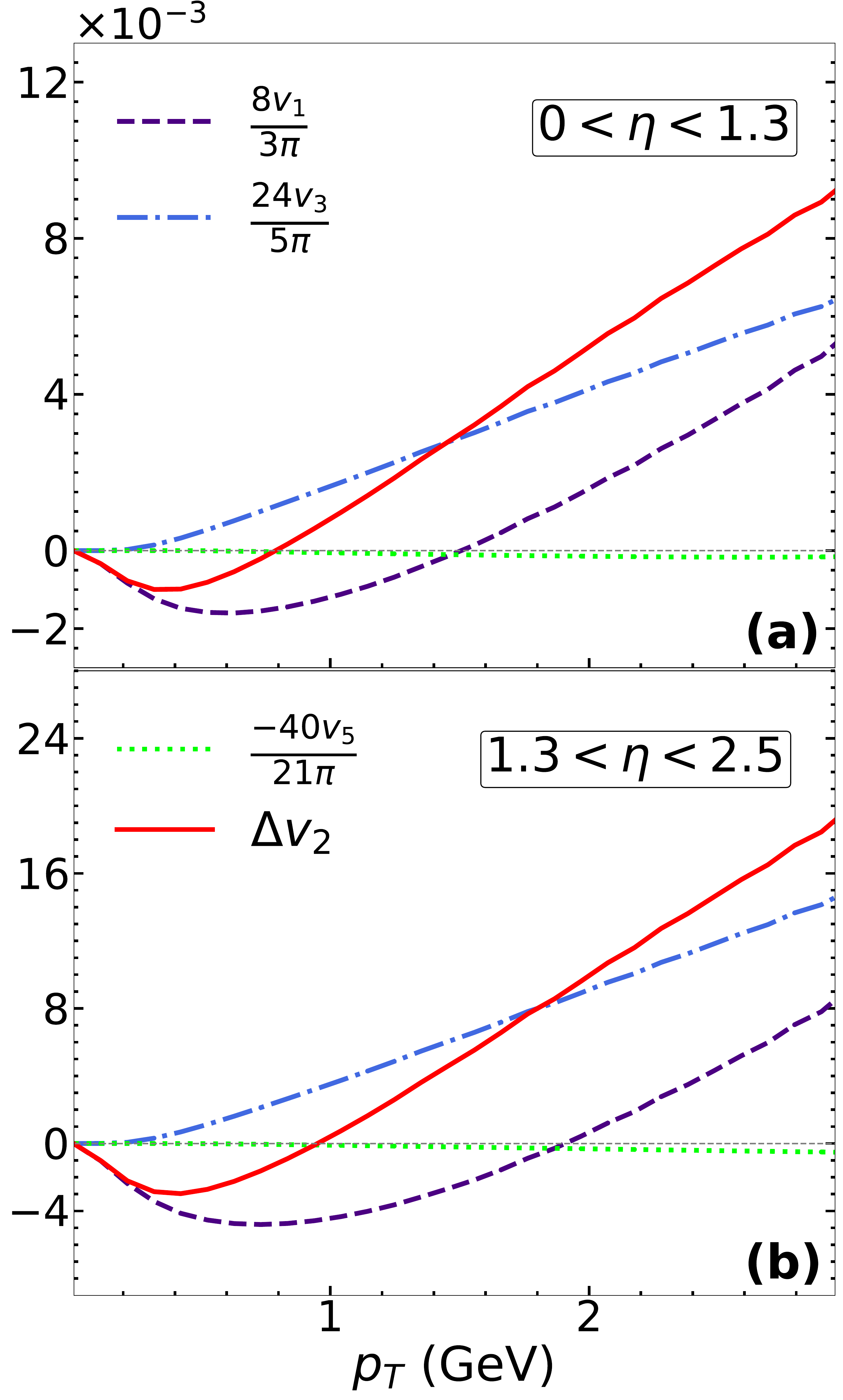}
 \caption{(Color online) The prediction for $\Delta v_2$ vs $p_T$ with tilt initial condition has been plotted in red solid line. 
 Further, the first three leading flow harmonics that contribute to $\Delta v_2$ (see Eq.~\ref{eq.deltav2ap}) are shown as well.}
 \label{fig.deltav2pT}
 \end{center}
\end{figure}

The $p_T$ dependence is shown in 
Fig.~\ref{fig.deltav2pT}. The predictions for $0<\eta<1.3$ and $1.3<\eta<2.5$ are shown separately in panels (a) and (b) 
respectively. 
As expected from the $\eta$ dependence, the magnitude of the split is larger in the $1.3<\eta<2.5$, otherwise, the 
trends are similar. $\Delta v_2$ is negative for small $p_T$. There is a turning point around $p_T\sim0.4$ GeV after 
which it crosses zero at $p_T\sim1$ GeV and monotonically rises. This $p_T$ dependence of $\Delta v_2$ is mainly 
borrowed from $v_1$ which also has a similar trend unlike $v_3$ which remains positive for all $p_T$. It is interesting to 
note that for $p_T>1.5$ GeV, $v_3$ becomes the dominant contributor.

\begin{figure}
 \begin{center}
 \includegraphics[scale=0.22]{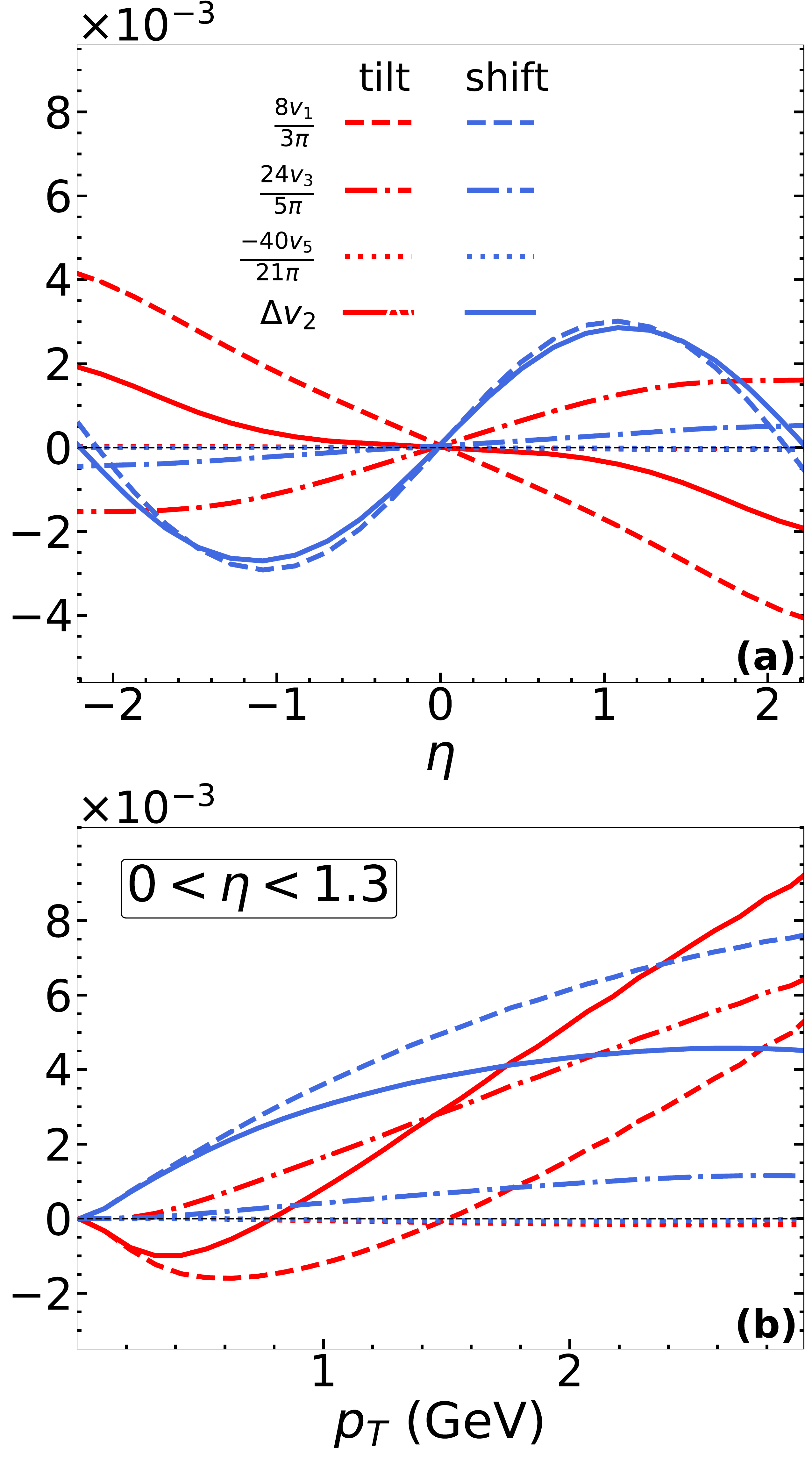}
 \caption{(Color online) The predictions for phase space dependence of $\Delta v_2$ has been compared between tilt initial 
 condition and shift initial condition. Further, the first three leading flow harmonics that contribute to $\Delta v_2$ are also shown 
 for each model to understand the origin in the difference of their phase space dependence of $\Delta v_2$. $\Delta v_2$ vs $\eta$ 
 is shown in panel (a) and $\Delta v_2$ vs $p_T$ is shown in panel (b).}
 \label{fig.compare}
 \end{center}
\end{figure}

Further, we present a comparative study of $\Delta v_2$ computed with TIC and SIC in Fig.~\ref{fig.compare}. 
The results on the $\eta$ dependence of $\Delta v_2$ are presented in panel (a) while the $p_T$ 
dependence is plotted in panel (b). As seen in panel (a), for $\vert \eta \vert < 1$, $\Delta v_2$ is around 10-20 times larger 
in SIC as compared to TIC. Further they are of opposite signs- at positive 
rapidities while the SIC gives a positive $\Delta v_2$, the TIC yields a negative $\Delta v_2$. This may be 
traced to the fact that $v_1$ is of opposite sign in the two models. Further, from panel (b) we note that the origin of this opposite 
sign is from the low $p_T$ region as for $p_T>1.5$ GeV, both models yield positive values. While we noted earlier that $\Delta v_2$ results out of tension between $v_1$ and $v_3$ in the TIC, in the SIC we find that it is 
mostly controlled by $v_1$ as $v_3$ turns out to be small in this case. It is worth noting here that an earlier study of $\Delta v_2$ 
within the transport framework of the AMPT model had similar conclusions as the results here with SIC~\cite{Zhang:2021cjt}. Thus, 
the characteristics of $\Delta v_2$ can serve as a sensitive probe of rapidity dependent initial condition of the fireball. 

\begin{figure}
 \begin{center}
 \includegraphics[scale=0.2]{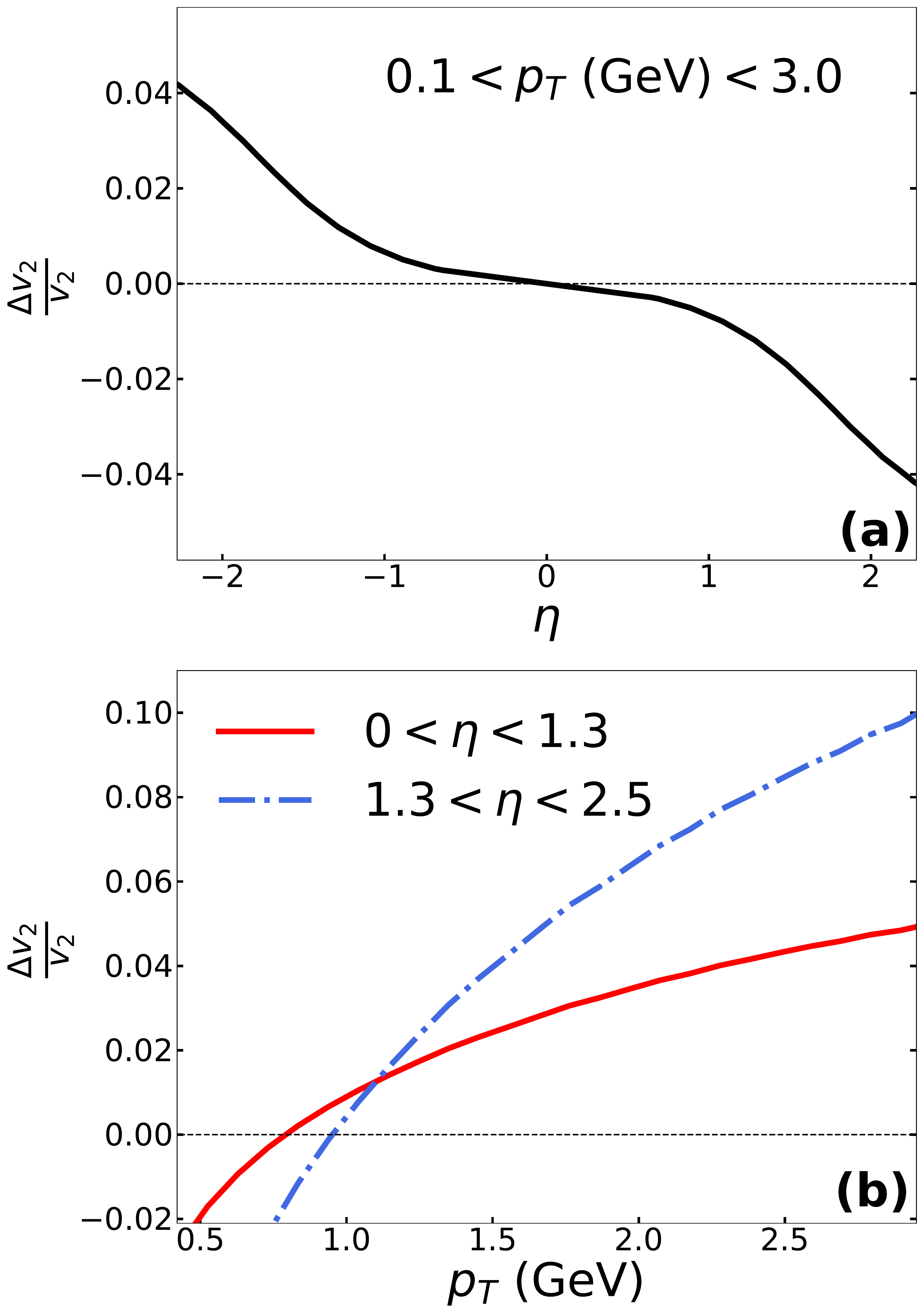}
 \caption{(Color online) The predictions for phase space dependence of $\Delta v_2 / v_2$ has been shown for tilt initial condition. 
 Panel (a) shows the $\eta$ dependence while panel (b) shows the $p_T$ dependence.}
 \label{fig.ratio}
 \end{center}
\end{figure}

It is well known that the various model systematics of the transverse initial condition can result in the variation of $v_2$~\cite{Shen:2011zc,Qiu:2011iv,Ruggieri:2013bda}. Such model dependencies can creep into our predictions of $\Delta v_2$ 
as well. We suggest to scale $\Delta v_2$ with $v_2$ in order to cancel out such systematics of the 
transverse initial condition which are not of interest here. We present the results of $\Delta v_2$ scaled by $v_2$ in Fig.~\ref{fig.ratio}.  
The $\eta$ dependence is plotted in panel (a). $\vert\Delta v_2/v_2\vert$ stays below 0.005 for $\vert\eta\vert<1$ beyond which it has 
a rapid linear growth. The $p_T$ differential values are plotted in panel (b). We obtain about $3\%$ and $5\%$ $\Delta v_2/v_2$ 
at $p_T\sim1.5$ GeV for $0<\eta<1.3$ and $1.3<\eta<2.5$ respectively. The ratio of phase space integrated $\Delta v_2$ and $v_2$, 
$\Delta v_2/ v_2$ comes out to be -0.0019 and -0.0268 for $0<\eta<1.3$ and $1.3<\eta<2.5$ respectively.

\section{SUMMARY}

The collision geometry of a non-central relativistic heavy ion collision introduces a large angular momentum in its initial state. 
Rapidity odd directed flow is a natural consequence of this. Such observables probe the longitudinal profile of the fireball. Recently, 
it has been observed that such large angular momentum in the initial state also causes a split in the magnitude of $v_2$, $\Delta v_2$ 
in different regions of the final state hadron momentum space - parallel and anti-parallel to the impact parameter direction~\cite{Chen:2021wiv,Zhang:2021cjt}. $\Delta v_2$ has been proposed to be a sensitive probe of the initial rapidity profile 
of the fireball. A transport model framework was adopted in these works that do not describe the $v_1$ data.

In this work, we have revisited the estimation of $\Delta v_2$ within a 3+1 D relativistic hydrodynamic framework with tilted initial 
condition that is known to describe the $v_1$ data~\cite{Bozek:2010bi}. We find $v_1$ to be the leading contributor to $\Delta v_2$, 
as was reported 
in earlier studies. However, unlike in those transport model based studies, we find that the tilted initial condition gives rise to 
sizeable rapidity odd $v_3$ which also contributes significantly to $\Delta v_2$, particularly for $p_T>1.5$ GeV, it becomes the 
dominant contributor to $\Delta v_2$. In order to demonstrate the sensitivity of $\Delta v_2$ 
to the choice of the initial condition, we have computed $\Delta v_2$ also with shifted initial condition. In this case, similar to the 
earlier studies, $v_3$ comes out to be negligible and hence $\Delta v_2$ gets contribution dominantly from $v_1$. The $\eta$ 
dependence of $\Delta v_2$ is of opposite sign for titled versus shifted initial conditions owing to the opposite signs of their $v_1$.
Finally, we have also presented the ratio of $\Delta v_2$ to $v_2$ so as to get rid of the various model uncertainties that 
affect $v_2$ and hence also $\Delta v_2$: for $p_T\sim1.5$ GeV we obtain $\Delta v_2/v_2\sim3\%$ and $5\%$ for $0<\eta<1.3$ 
and $1.3<\eta<2.5$ respectively. Our study demonstrates that $\Delta v_2$ can play complementary role to $v_1$ 
in constraining the rapidity profile of the fireball.

\section{ACKNOWLEDGEMENTS}
SC acknowledges helpful discussions with Piotr Bozek and IISER Berhampur for Seed Grant.

\bibliographystyle{apsrev4-1}
\bibliography{Splitv2}

\end{document}